\pdfoutput=1

\documentclass[sigconf]{acmart}

\usepackage{algorithm}
\usepackage{algpseudocode}
\usepackage{float}
\restylefloat{algorithm}
\usepackage{placeins}
\usepackage{etoolbox}
\usepackage{tikz}
\usetikzlibrary{shapes.geometric, arrows.meta, positioning, fit,
                backgrounds, calc, decorations.pathreplacing}
\usepackage{standalone}
\usepackage{adjustbox}
\usepackage{enumitem}
\newcommand{\sidefontT}{\sffamily\bfseries\fontsize{7}{8}\selectfont}
\newcommand{\sidefontL}{\sffamily\fontsize{6.5}{7.5}\selectfont}
\newcommand{\sidefontS}{\sffamily\fontsize{5.5}{6.5}\selectfont}

\newcommand{\sidebrain}{%
\begin{tikzpicture}[x=1mm,y=1mm,line cap=round,line join=round,
    sulcus/.style={black!38,line width=.4pt},
    ridge/.style={white,line width=.35pt,opacity=.8},
    flow/.style={-{Stealth[length=1.1mm,width=.9mm]},line width=.45pt,draw=black!70}]
  \path (0,21) -- (22,21);                         
  \node[font=\sidefontT,text=black!60] at (11,17.5) {Brain};
  \fill[black!9] (11.5,-7.3) ellipse (8.5 and .9);
  \shade[inner color=red!8,outer color=red!38]
    plot[smooth cycle,tension=.7] coordinates
    {(14.2,0.6) (17.5,1.0) (20.4,-0.2) (20.2,-2.6) (17.2,-3.6) (14.4,-2.4)};
  \draw[black!55,line width=.5pt]
    plot[smooth cycle,tension=.7] coordinates
    {(14.2,0.6) (17.5,1.0) (20.4,-0.2) (20.2,-2.6) (17.2,-3.6) (14.4,-2.4)};
  \foreach \yy in {-0.6,-1.6,-2.5}{%
    \draw[black!35,line width=.3pt]
      (15,\yy) .. controls (17,\yy+.5) and (18.5,\yy+.3) .. (19.8,\yy-.2);}
  \shade[left color=red!30,right color=red!8]
    (11.6,0.4) .. controls (12.2,-2.5) and (12.1,-4.5) .. (12.6,-6.4)
    -- (14.2,-6.4) .. controls (14.1,-4.4) and (14.3,-2.4) .. (14.6,0.2) -- cycle;
  \draw[black!55,line width=.5pt]
    (11.6,0.4) .. controls (12.2,-2.5) and (12.1,-4.5) .. (12.6,-6.4)
    -- (14.2,-6.4) .. controls (14.1,-4.4) and (14.3,-2.4) .. (14.6,0.2);
  \shade[inner color=red!4,outer color=red!34]
    plot[smooth cycle,tension=.75] coordinates
    {(0.3,4.5) (1.5,8.5) (5,11.3) (9.5,12.3) (14.5,11.7) (18.8,9.6)
     (21.6,6.2) (21.8,3) (19.6,1.2) (15.5,0.6) (12.5,-0.6) (8.5,-0.3)
     (4.5,0) (1.6,1.8)};
  \fill[white,opacity=.45,rotate around={12:(7.5,9)}] (7.5,9) ellipse (4.8 and 1.6);
  \draw[black!60,line width=.6pt]
    plot[smooth cycle,tension=.75] coordinates
    {(0.3,4.5) (1.5,8.5) (5,11.3) (9.5,12.3) (14.5,11.7) (18.8,9.6)
     (21.6,6.2) (21.8,3) (19.6,1.2) (15.5,0.6) (12.5,-0.6) (8.5,-0.3)
     (4.5,0) (1.6,1.8)};
  \foreach \p in {%
      {(5.5,4.2) .. controls (9,5) and (12,5.4) .. (15,7)},
      {(11.2,12.1) .. controls (10.2,10) and (12,8.4) .. (11,6.2)},
      {(2.5,7) .. controls (4,8.2) and (5.5,6.8) .. (7.5,8.3)},
      {(3.8,10) .. controls (5.5,9.6) and (6.5,10.8) .. (8.5,10.3)},
      {(14,10.8) .. controls (15,9) and (16.5,9.6) .. (17.5,8)},
      {(15.5,5.2) .. controls (17,6.2) and (18.5,4.4) .. (20.5,5.3)}}{%
    \draw[sulcus] \p;
    \draw[ridge,yshift=.3mm] \p;}
  \draw[orange!30,line width=3.2pt]
    (7.6,1.6) .. controls (9.5,0.9) and (11.8,1.4) .. (12.8,2.6)
    .. controls (13.6,3.6) and (12.8,4.4) .. (12.1,3.7);
  \draw[orange!80!black,line width=1.5pt]
    (7.6,1.6) .. controls (9.5,0.9) and (11.8,1.4) .. (12.8,2.6)
    .. controls (13.6,3.6) and (12.8,4.4) .. (12.1,3.7);
  \fill[orange!80!black] (7.6,1.6) circle (.9);
  \draw[orange!70!black,densely dashed,line width=.4pt] (10.2,2.5) circle (3.6);
  \draw[black!30,line width=.35pt] (6.8,1.4) -- (0.4,-12);
  \draw[black!30,line width=.35pt] (13.6,1.4) -- (21.6,-12);

  \draw[fill=white,draw=black!45,line width=.5pt,rounded corners=1.2pt]
    (0,-12) rectangle (22,-41);
  \node[font=\sidefontL\bfseries,text=orange!70!black] at (11,-14) {Hippocampus};
  \node[font=\sidefontS,fill=black!10,draw=black!45,line width=.35pt,
        rounded corners=.8pt,minimum width=11mm,minimum height=3mm,inner sep=.4mm]
        (cx) at (11,-18.2) {Neocortex};
  \node[font=\sidefontS\bfseries,text=white,fill=blue!72!black,rounded corners=.8pt,
        minimum width=6.5mm,minimum height=3.4mm,inner sep=.3mm] (ec) at (5,-36) {EC};
  \draw[orange!68!black,line width=2.2pt]
    (12,-29) ++(170:6) arc[start angle=170,end angle=80,radius=6];
  \draw[green!58!black,line width=2.2pt]
    (12,-29) ++(80:6) arc[start angle=80,end angle=-70,radius=6]
    .. controls (13.2,-35.6) and (11.4,-33.4) .. (12.2,-31.6);
  \draw[blue!45,line width=1.8pt]
    (12.3,-31.5) ++(20:2.6) arc[start angle=20,end angle=260,radius=2.6];
  \node[font=\sidefontS\bfseries,text=orange!68!black] at (7.1,-22.0) {CA1};
  \node[font=\sidefontS\bfseries,text=green!50!black]  at (19.3,-24.8) {CA3};
  \node[font=\sidefontS\bfseries,text=blue!60!black]   at (8.4,-31.3) {DG};
  \draw[{Stealth[length=1.1mm,width=.9mm]}-{Stealth[length=1.1mm,width=.9mm]},
        line width=.45pt,draw=black!70]
    (cx.west) -| (2.6,-34.2);
  \draw[flow] (ec.east) .. controls (9.6,-36) and (9.8,-35.2) .. (10.1,-34.1);
  \draw[flow] (6.0,-28.6) -- (5.6,-34.1);
  \node[font=\sidefontS,text=black!55] at (11,-39.3)
    {EC$\to$DG$\to$CA3$\to$CA1$\to$EC};
\end{tikzpicture}}

\newcommand{\sidefpga}{%
\begin{tikzpicture}[x=1mm,y=1mm]
  \path (-1.2,21) -- (21.2,21);                    
  \node[font=\sidefontT,text=black!60] at (10,17.5) {FPGA};
  \foreach \i in {0,...,8}{%
    \fill[black!45] ({1.6+\i*2.1},13)   rectangle ++(0.9,1.1);
    \fill[black!45] ({1.6+\i*2.1},-7)   rectangle ++(0.9,-1.1);
    \fill[black!45] (0,{-5.4+\i*2.1})   rectangle ++(-1.1,0.9);
    \fill[black!45] (20,{-5.4+\i*2.1})  rectangle ++(1.1,0.9);}
  \draw[fill=black!78,draw=black,rounded corners=.8pt] (0,-7) rectangle (20,13);
  \draw[fill=black!8,draw=black!40,line width=.3pt] (2.5,-4.5) rectangle (17.5,10.5);
  \foreach \cx in {0,...,5}{\foreach \cy in {0,...,5}{%
    \fill[black!22] ({3.3+\cx*2.4},{-3.7+\cy*2.4}) rectangle ++(1.8,1.8);}}
  \fill[black!55] (3.1,-3.9) rectangle (5.3,10.3);
  \node[font=\sidefontS\bfseries,text=white,rotate=90] at (4.2,3.2) {PS};
  \foreach \cy/\cc in {5/blue!72!black,4/blue!72!black,3/green!58!black,
                       2/green!58!black,1/orange!68!black,0/orange!68!black}{%
    \foreach \cx in {1,...,5}{%
      \fill[\cc] ({3.3+\cx*2.4},{-3.7+\cy*2.4}) rectangle ++(1.8,1.8);}}
  \node[font=\sidefontS\bfseries,text=white,fill=blue!72!black,inner sep=.3mm]   at (11.4,8.0)  {DG/EC};
  \node[font=\sidefontS\bfseries,text=white,fill=green!58!black,inner sep=.3mm]  at (11.4,3.2)  {CA3};
  \node[font=\sidefontS\bfseries,text=white,fill=orange!68!black,inner sep=.3mm] at (11.4,-1.6) {CA1};
  \node[font=\sidefontL\bfseries,text=black!70,anchor=north] at (10,-9.8) {Zynq US+};
  \node[font=\sidefontS,text=black!55,align=center,anchor=north] at (10,-12.6)
    {XCZU9EG\\PS + PL fabric};
\end{tikzpicture}}

\makeatletter
\def\@secfont{\normalfont\bfseries\fontsize{12}{13.5}\selectfont\section@raggedright}
\def\@subsecfont{\normalfont\bfseries\fontsize{10}{11.5}\selectfont\section@raggedright}
\def\@subsubsecfont{\normalfont\itshape\fontsize{9}{10.5}\selectfont}
\makeatother
\setcopyright{none}
\renewcommand\footnotetextcopyrightpermission[1]{}

\begin{document}
\raggedbottom

\title{HippocamFLEX: FPGA Prototyping of Hippocampus-Inspired
Architecture for Context-Aware Sequence Prediction}

\author{Bhuvanesh S. Kayarabettu}
\affiliation{%
  \institution{University of Wisconsin--Madison}
  \city{Madison}
  \state{WI}
  \country{USA}
}
\email{kayarabettu@wisc.edu}

\author{Hamed Poursiami}
\affiliation{%
  \institution{George Mason University}
  \city{Fairfax}
  \state{VA}
  \country{USA}
}
\email{hpoursia@gmu.edu}

\author{Keiland Cooper}
\affiliation{%
  \institution{University of California, Irvine}
  \city{Irvine}
  \state{CA}
  \country{USA}
}
\email{kwcooper@uci.edu}

\author{Babak Shahbaba}
\affiliation{%
  \institution{University of California, Irvine}
  \city{Irvine}
  \state{CA}
  \country{USA}
}
\email{babaks@uci.edu}

\author{Norbert Fortin}
\affiliation{%
  \institution{University of California, Irvine}
  \city{Irvine}
  \state{CA}
  \country{USA}
}
\email{norbert.fortin@uci.edu}

\author{Maryam Parsa}
\affiliation{%
  \institution{George Mason University}
  \city{Fairfax}
  \state{VA}
  \country{USA}
}
\email{mparsa@gmu.edu}

\author{Akhilesh Jaiswal}
\affiliation{%
  \institution{University of Wisconsin--Madison}
  \city{Madison}
  \state{WI}
  \country{USA}
}
\email{akhilesh.jaiswal@wisc.edu}

\renewcommand{\shortauthors}{Kayarabettu, Jaiswal et al.}

\begin{abstract}
Unlike conventional architectures that treat memory as static pattern storage,
the hippocampus organizes experience as ordered sequences and retrieves them through
temporally compressed replay spanning prior, current, and anticipated states. This capacity
of context-appropriate sequence prediction remains absent from conventional deep learning accelerators
and existing neuromorphic fabrics. We present \textit{HippocamFLEX}, a fully synthesizable
digital implementation of these hippocampal computational principles, realized with
SystemVerilog RTL on the AMD Zynq UltraScale+,XCZU9EG. The design maps the hippocampal
trisynaptic pathway onto three pipeline stages: a Dentate Gyrus / Entorhinal Cortex (DG/EC)
stage for hypervector proximity matching, a CA3 stage for context-gated recurrent sequence
prediction with multi-phase temporal tagging, and a CA1 stage that integrates past, present,
and predicted future events into a theta-sequence inspired output. While much brain-inspired
hardware relies on analog mixed-signal design for energy efficiency, this work targets
FPGA prototyping as a deliberate architectural choice: FPGAs offer rapid
design-space exploration without tape-out cost, bit-exact determinism that simplifies
verification. The design supports 64 events across
 16 contexts and is verified on physical silicon across context-isolation, temporal-ordering,
 and mid-sequence sliding-window test cases and is inherently scalable to larger event and context sets. 
\end{abstract}



\maketitle
\fancypagestyle{generic}{%
  \fancyhf{}%
  \fancyhead[L]{\footnotesize HippocamFLEX: FPGA Prototyping of Hippocampus-Inspired Architecture}%
  \fancyhead[R]{\footnotesize \shortauthors}%
  \fancyfoot[C]{\footnotesize\thepage}%
  \renewcommand{\headrulewidth}{0pt}}
\pagestyle{generic}

\section{Introduction}
\label{sec:intro}

The hippocampus encodes, organises, and replays sequences of events
in context~\cite{eichenbaum2004,allen2013}.
It must simultaneously separate overlapping inputs, associate events
with their context, predict upcoming events, and compress that
past-present-future trajectory into a single oscillatory cycle.
Electrophysiology has mapped these functions onto four sub-regions
(Fig.~\ref{fig:neuro_overview}): the Entorhinal Cortex (EC) provides
compact event and context codes; the Dentate Gyrus (DG)
orthogonalises overlapping patterns; CA3 stores sequence associations
via recurrent collaterals; and CA1 integrates CA3 predictions with a
decaying EC trace into temporally-ordered theta inspired
sequences~\cite{shahbaba2022}.

Existing brain-inspired and AI hardware fail to deliver key architectural features required for efficient context-appropriate sequence prediction. Deep learning accelerators~\cite{chen2014,cavigelli2015} lack
temporal context; neuromorphic fabrics~\cite{merolla2014,davies2018}
implement spike dynamics but not the hippocampal
pipeline~\cite{schuman2022}inspired by the unique structure and organization of hippocampal sub-regions.

Notably, context-aware sequence prediction requires more than retrieving a similar
stored event. The same event can occur in several contexts with different
successors, so a correct recall must identify the context, restrict the
associated events, and recover their temporal order. At a sequence terminus,
the system must also distinguish the absence of a continuation from a failed
lookup. 
HippocamFLEX implements these operations as a parameterized digital pipeline.
Parallel event and context searches feed a context-gated recurrent crossbar;
stored timing codes distinguish the present event from its two predicted
successors. CA1 combines those predictions with a decaying trace of observed
events and assigns an explicit rank to each active event. The FPGA
implementation exposes memory capacities, matching thresholds, and recurrent
depth as synthesis parameters, while the AXI4-Stream interface provides a
single path for loading memories and collecting recall results.

\textbf{HippocamFLEX}, on the Zynq UltraScale+ XCZU9EG, contributes:
(1)~a complete three-stage digital pipeline (DG/EC$\to$\allowbreak CA3$\to$\allowbreak CA1)
in eleven SystemVerilog modules at 100\,MHz;
(2)~a seven-opcode AXI4-Stream ISA enabling memorisation and recall
over AXI DMA; and
(3)~a bare-metal AArch64 driver verified on silicon across
context-isolation, temporal-ordering, and sequence-terminus cases.
The evaluation covers both functional behavior and implementation cost.

\section{Related Work}
\label{sec:related}

\subsection{Memory Models and Hardware Architectures}
MFEC~\cite{blundell2016} and NEC~\cite{pritzel2017} demonstrated that
similarity-based episodic caching accelerates reinforcement learning;
MERLIN~\cite{wayne2018} added predictive structure;
HAMI~\cite{poursiami2025} employs context-sensitive indexing closest
to hippocampal function. However, all of the above works are software-only implementation with no description of a pathway to
hardware emulation.

TrueNorth~\cite{merolla2014} and Loihi~\cite{davies2018} scale
spiking dynamics to millions of neurons; Neurogrid~\cite{benjamin2014}
pursues mixed signal analog emulation. For sequence prediction, Kreiser
et~al.~\cite{kreiser2018} used dynamic neural fields, Bouhadjar
et~al.~\cite{bouhadjar2023} coupled spiking neural networks (SNNs) with resistive RAM, and Siegel
et~al.~\cite{siegel2023} achieved context-aware prediction with
memristive neurons. None of the above works assembles EC matching, CA3 context-gated
prediction, and CA1 integration in one end-to-end validated
system~\cite{schuman2022}.

Further, feed-forward deep learning accelerators~\cite{chen2014,cavigelli2015} have no
associative recall. Hopfield networks~\cite{hopfield1982,hopfield1984}
and dense associative memories~\cite{krotov2016,ramsauer2021} provide
content-addressable recall but lack context-gating mechanism.
Memristor~\cite{li2015} and analog~\cite{stamant2014} accelerators
trade reconfigurability for energy efficiency.

\subsection{HippocamPEX: Mixed-Signal Hippocampal Architecture}
HippocamPEX~\cite{hippocampex2025} develops an end-to-end mixed-signal
architecture for context-aware sequence prediction. Its EC module uses
6T-SRAM+4T-CAM (content addressable memory) arrays; CA3 combines a context-gated
heteroassociative crossbar with a multi-port CAM sequence predictor; and
CA1 integrates capacitor-based past traces with present and future event
signals through analog ramp-based temporal ordering. HippocamPEX connects these
circuits to a neuroscience model and evaluates their representational
dynamics and context-specific sequence predictions through analog circuit
simulations in GlobalFoundries 22\,nm FDX and a MATLAB hardware model.
The present work - HippocamFLEX - addresses the complementary FPGA implementation of
context-gated sequence prediction, with parameterized RTL, system-level
integration, and on-silicon validation.

\section{Methodology}
\label{sec:methodology}

\subsection{Neuroscience Abstraction}
\label{sec:neuro}

\begin{figure*}[!t]
\centering
\makebox[\textwidth][c]{%
  \adjustbox{valign=t}{\sidebrain}\hspace{3mm}%
  \adjustbox{valign=t}{\includegraphics[height=6.5in]{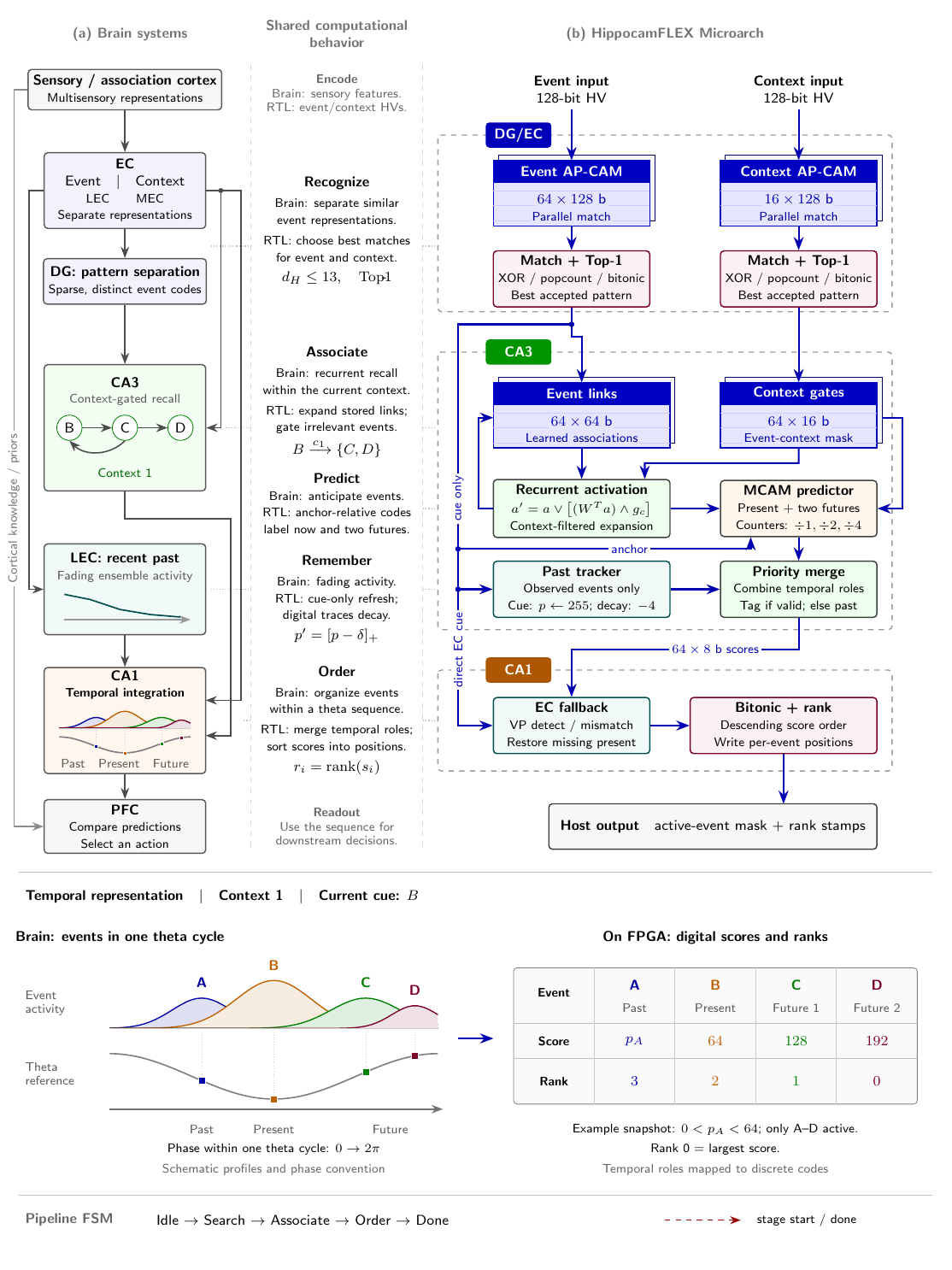}}%
  \hspace{3mm}\adjustbox{valign=t}{\sidefpga}}
\caption{Mapping hippocampal computations to HippocamFLEX.
  \textbf{Left:} Biological regions; \textbf{centre:} shared computations;
  \textbf{right:} hardware blocks. The timeline compares biological event
  activity with FPGA scores and temporal ranks.}
\Description{Three-column diagram mapping hippocampal sub-regions
  to shared computations and HippocamFLEX RTL hardware blocks}
\label{fig:neuro_overview}
\end{figure*}

Figure~\ref{fig:neuro_overview} maps hippocampal sub-regions to HippocamFLEX hardware;
four neuroscience principles guide the design.

\textbf{Functional segregation.}
The EC encodes rapidly-changing events (LEC) and slowly-drifting context
(MEC) as separate distributed representations~\cite{eichenbaum2004}.
In HippocamFLEX this becomes two independent Proximity-CAM banks: one for up to
64 event hypervectors and one for 16 context hypervectors, searched in
parallel on every recall.

\textbf{Pattern separation.}
The DG orthogonalises overlapping EC inputs through massive cell-count
expansion and sparse activity. In hardware: a Hamming proximity threshold suppresses any stored hypervector whose mismatch with the query exceeds 13 bits ($\approx$10\% of 128), admitting only matches at least 89.8\% similar.

\textbf{Context-gated sequence coding.}
CA3's recurrent collaterals store sequence associations through Hebbian
plasticity, and its conjunction of DG event inputs with EC context
inputs at distal dendrites enables context-gated recall.
This maps to a 64$\times$64 heteroassociative crossbar gated by a
64$\times$16 context-gate memory, iterated for 7 recurrent hops.

\textbf{Theta-sequence integration.}
CA1 receives CA3 predictions (poly-synaptic path) and a decaying EC trace
(mono-synaptic path) simultaneously, producing temporally-ordered theta
sequences~\cite{shahbaba2022}. In hardware CA1 is given both paths
explicitly: the CA3 merged voltage array (poly-synaptic) and the DG/EC
top-1 event one-hot (mono-synaptic). All 64 event voltage registers are
ranked by a pipelined bitonic sorting network, and the resulting rank
positions form the theta-sequence output.

\subsection{FPGA as a Prototyping Platform}
\label{sec:fpga}

We target the AMD Zynq UltraScale+ XCZU9EG (ZCU102).
On an FPGA, all seven memory depths and every numerical parameter
(\texttt{MAX\_MISMATCH}, \texttt{DECAY\_STEP}, counter divisors) are
Verilog parameters re-synthesised in minutes, enabling rapid
architecture sweeps and bit-exact simulation-to-silicon traceability.
The ZCU102's Cortex-A53 PS handles memorisation and result collection;
the PL executes the hippocampal recall pipeline at 100\,MHz.

\subsection{Hyper-dimensional Encoding and the Proximity-CAM Abstraction}
\label{sec:hdc}

Events and contexts use 128-bit binary \emph{hypervectors} (HVs).
Independent random vectors share approximately 50\% of their bits;
the benchmark in Section~\ref{sec:results} instead composes correlated
digit and colour representations. As a dimensionality comparison,
VS-Graph~\cite{poursiami2025vsgraph} reports less than 1.5\% accuracy
degradation when reducing HV width from 8192 to 128 in graph
classification. That result concerns a different workload.

\textbf{Proximity-CAM lookup.}
The core lookup primitive in hyper-dimensional systems is
nearest-prototype retrieval: given a query vector, find the stored
vector with highest similarity.
In software frameworks such as VS-Graph~\cite{poursiami2025vsgraph},
this reduces to a cosine-similarity scan over class prototypes.
The \emph{Proximity Content-Addressable Memory} (Proximity-CAM) searches
stored hypervectors by similarity rather than by an explicit address.
HippocamFLEX computes the Hamming distance
$d[i] \gets \$\mathrm{countones}(q \oplus m_i)$,
computed for all stored entries in parallel in one registered pipeline
stage.
The Proximity-CAM banks use flip-flop register files to make every stored
row available to the parallel comparison logic in that stage. In the
memory organization used here,
a BRAM's mandatory output register would push the result to a second
stage, breaking the pipeline handshake.
The 8-bit distance represents all mismatch counts from 0 to 128;
the threshold gate admits distances up to 13 and suppresses the rest.

\section{HippocamFLEX Architecture and Algorithm}
\label{sec:arch-alg}

HippocamFLEX operates in two phases: \emph{memorisation} and \emph{recall}. During memorisation, hypervectors
(compact binary codes for events and contexts) are written into the
on-chip memory banks (Proximity-CAMs), sequence associations are written to
the crossbar, and timing tags are written to the multi-port
content-addressable memory (MCAM). The MCAM stores a temporal code for
each event--context pair and compares it with three counter values to
identify the present event and its next two successors.
During \emph{recall} (retrieval), a query pair triggers pattern completion in CA3 and
theta inspired-sequence replay in CA1, returning the predicted next events with
timestamps.
Algorithm~\ref{alg:hippocamflex} gives the complete sequence for both phases.

 \textbf{Microarchitecture structure.}
The three pipeline stages are implemented in SystemVerilog-HDL and sequenced
by a six-state Pipeline Controller FSM that fires a one-cycle \emph{start}
pulse to each stage in turn, waits for its one-cycle \emph{done} reply, and
advances.
No stage begins before the previous one has finished.
Memorisation commands enter through a mutually exclusive TRAIN state,
active only when the recall path is idle.
Table~\ref{tab:hierarchy} has all sub-modules.

\begin{table}[t]
  \caption{Notation used in Algorithm~\ref{alg:hippocamflex}.}
  \label{tab:notation}
  \footnotesize
  \begin{tabular}{@{}l@{\;=\;}l@{}}
    \toprule
    $C_k$, $E_j$   & context / event hypervectors ($k{=}0\ldots15$, $j{=}0\ldots63$)\\
    $\mathcal{A}$  & associations: $(j_f\!\to\!j_t)$ means $E_{j_f}$ precedes $E_{j_t}$\\
    $\mathcal{G}$  & gates: $(j,k)$ means event $j$ belongs to context $k$\\
    $\mathcal{T}$  & timing codes: $(j,k,t)$ is position of $E_j$ in context $C_k$\\
    VP/VF1/VF2     & voltage tags 0x40 / 0x80 / 0xC0 (present / next / two-ahead)\\
    $q \oplus m$   & bitwise XOR;\; $\$\mathrm{countones}(\cdot)$ = popcount\\
    \bottomrule
  \end{tabular}
\end{table}

\begin{algorithm}[t]
\caption{HippocamFLEX: Memorisation and Recall}
\label{alg:hippocamflex}
\small
\begin{algorithmic}[1]
\Statex \textbf{Phase 1: Memorisation}
\State \textbf{Init:} Reset chip via GPIO; initialise AXI DMA; release reset.
\State \textbf{CMD\,0x1--0x2}: Write each $C_k$ and $E_j$ into their
       Proximity-CAM slots.
\State \textbf{CMD\,0x3--0x4}: For each $(j_f\!\to\!j_t)\!\in\!\mathcal{A}$
       and $(j,k)\!\in\!\mathcal{G}$, set crossbar and context-gate bits.
\State \textbf{CMD\,0x5}: For each $(j,k,t)\!\in\!\mathcal{T}$,
       write timing code $t$ into $\mathrm{MCAM}[j][k]$.
\Statex \textbf{Phase 2: Recall}
\State \textbf{CMD\,0x6}: Send query $(C_q, E_q)$ as 9-beat AXI packet.
\Statex \textit{Stage~1 -- DG/EC: Pattern Matching (25 cycles, fixed)}
\State \textbf{Parallel} for all $i,j$:\;
       $d_i\!\gets\!\$\mathrm{countones}(C_q\!\oplus\!C_i)$;\;
       $v_i\!\gets\!(d_i\!\le\!13)\,?\,(255\!-\!d_i)\!:\!0$.\;
       Same for event bank using $E_q$.
\State Bitonic rank (12\,cyc context / 23\,cyc event): latch Top-1 context $k^*$ and Top-1 event $j^*$.
\Statex \textit{Stage~2 -- CA3: Sequence Prediction ($t_\mathrm{ref}{+}24$ cycles; 27 at $t_\mathrm{ref}{=}3$)}
\State Register gate row for $k^*$;\; active $\gets \{j^*\}$.
\For{step $= 1\ldots7$} active $\gets$ active $\cup$ (context-gated neighbours).
\EndFor
\Statex \hfill\textit{(9 cycles start-to-done: 7 accumulating hops $+$ init $+$ done)}
\State CNTR-1 ($\div$1) sweeps from 0 until it matches $\mathrm{MCAM}[j^*][k^*]$:
       VP\,=\,0x40;\; lock $t_\mathrm{ref}$. \emph{($t_\mathrm{ref}{+}1$ cycles.)}
\State CNTR-2 ($\div$2) matches $t_\mathrm{ref}{+}1$: VF1\,=\,0x80 on next event.
\State CNTR-3 ($\div$4) matches $t_\mathrm{ref}{+}2$: VF2\,=\,0xC0 on two-ahead event.
       \emph{(9 cycles when both are found; 50-cycle sweep at a terminus.)}
\State Decay all non-anchor voltages by 4\,LSB/cycle;\; merge MCAM tags (priority)
       with decaying past voltage scores.
\Statex \textit{Stage~3 -- CA1: Temporal Integration (24 cycles)}
\State If no VP tag survives CA3, inject the mono-synaptic DG/EC event at VP.
\State Bitonic full rank over merged voltage scores; rank index becomes the
       timestamp. VF2 $\to$ ts\,=\,0 (most future);\;
       inactive $\to$ ts\,=\,0x3F.
\State \textbf{Return} $\mathit{voltages}[63{:}0]$, $\mathit{timestamps}[63{:}0]$, $\mathit{onehot}[63{:}0]$ via a 35-beat \texttt{M\_AXIS} response;\; assert IRQ.
\Statex \hfill\textit{Total: $31 + 2S + t_\mathrm{ref}$ cycles ($S{=}21$ bitonic stages at $E{=}64$) $=$ 76 at $t_\mathrm{ref}{=}3$.}
\end{algorithmic}
\end{algorithm}

\begin{table}[t]
  \caption{HippocamFLEX module hierarchy.}
  \label{tab:hierarchy}
  \footnotesize
  \resizebox{\columnwidth}{!}{%
  \begin{tabular}{ll}
    \toprule
    \textbf{Module} & \textbf{Function} \\
    \midrule
    \texttt{hippocamFLEX\_axis\_wrapper} & AXI4-S PS bridge, TX/RX FSMs \\
    \quad\texttt{hippocamFLEX\_top} & Core processor \\
    \quad\quad\texttt{control\_fsm} & 6-state pipeline sequencer \\
    \quad\quad\texttt{dg\_ec\_stage} & Stage~1: pattern recognition \\
    \quad\quad\quad\texttt{ap\_cam} $\times$2 & Context (16) \& event (64) CAMs \\
    \quad\quad\quad\texttt{sort\_comparator} $\times$2 & Top-1 bitonic rank \\
    \quad\quad\texttt{ca3\_stage} & Stage~2: Sequence prediction \\
    \quad\quad\quad\texttt{recurrent\_hetero\_crossbar} & 64$\times$64 assoc + 64$\times$16 gate \\
    \quad\quad\quad\texttt{mcam\_sequence\_predictor} & 3-counter VP/VF1/VF2 tagger \\
    \quad\quad\quad\texttt{temporal\_context\_tracker} & Per-event decay model \\
    \quad\quad\texttt{ca1\_stage} & Stage~3: Temporal integration \\
    \quad\quad\quad\texttt{sort\_comparator} & Full sort, TOP\_K\,=\,64 \\
    \bottomrule
  \end{tabular}%
  }
\end{table}

\section{DG/EC Stage: Pattern Recognition}
\label{sec:dgec}

\begin{figure*}[!t]
\centering
\adjustbox{max width=\textwidth, max totalheight=0.42\textheight, keepaspectratio}{\input{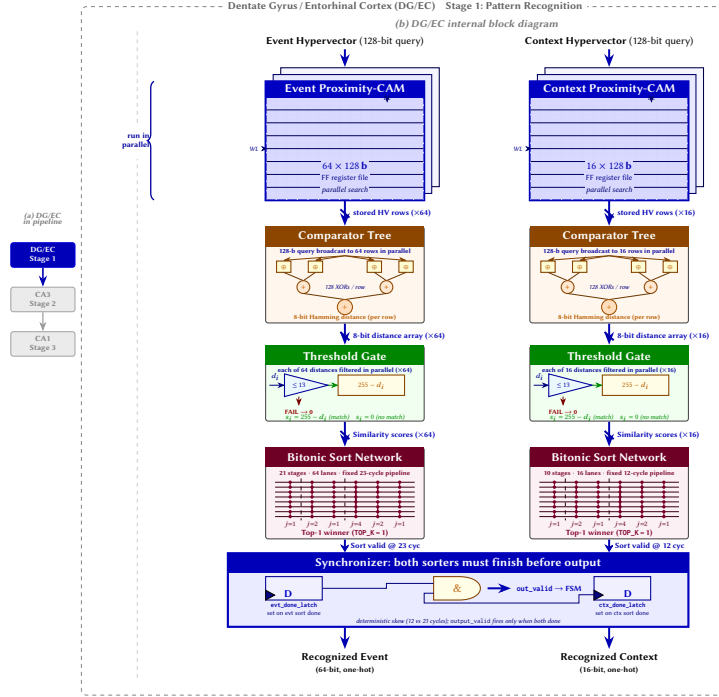}}

\caption{DG/EC Stage.
  \textbf{(a)}~Locator.
  \textbf{(b)}~Two Proximity-CAM banks (Event: 64$\times$128\,b; Context:
  16$\times$128\,b register files) search every stored row in parallel.
  Comparator Trees compute 8-bit Hamming distances;
  the Threshold Gate zeroes mismatches $>$13 and converts valid distances
  to voltage scores ($v{=}255{-}d$).
  Two pipelined bitonic sorting networks run in parallel (23 and 12
  cycles); the done-latch Synchronizer holds both results until the
  deeper network finishes, then asserts \texttt{out\_valid}.}
\Description{Block diagram of the DG/EC stage showing two Proximity-CAM
  banks, comparator trees, threshold gates, bitonic sorting networks and a
  done-latch synchronizer.}
\label{fig:dgec_arch}
\end{figure*}

The DG/EC stage finds the closest stored pattern in two parallel register-file banks--one for the
query event, one for the query context--and returns each as a one-hot index.
Figure~\ref{fig:dgec_arch} details the sub-blocks executed in order.

\textbf{Event and Context Proximity-CAMs (Fig.~\ref{fig:dgec_arch}, top).}
Each Proximity-CAM is a synchronous register file that, on a
single registered clock edge, compares a 128-bit query against
\emph{every} stored row simultaneously--not one at a time.
The Event bank (64$\times$128\,b) and Context bank (16$\times$128\,b)
are both searched on the same cycle and run fully in parallel,
as shown by the ``run in parallel'' brace in the figure.
Ruling out BRAM is mandatory: its output register would push the
result to a second pipeline stage, breaking the handshake with the
downstream sorter (see~\S\ref{sec:hdc}).
Distributed LUTRAM would satisfy that constraint at lower area, but
each array is cleared wholesale on reset and distributed RAM has no
reset port, so EDA tool infers registers instead; \S\ref{sec:implementation}
quantifies the cost.

\textbf{Comparator Trees.}
Each Comparator Tree broadcasts the 128-bit query to all stored
rows simultaneously and computes, for each row $i$:
$d[i]\!\gets\!\$\mathrm{countones}(q\!\oplus\!m_i)$.
Vivado synthesises each popcount as a 7-level binary adder tree
(128 XOR bits summed pairwise through carry-save and ripple-carry
stages), producing one 8-bit Hamming distance per row, registered
in one pipeline stage.

\textbf{Threshold Gate and Voltage Score Conversion.}
The Threshold Gate applies the DG's 90\% similarity criterion
($13/128\approx10\%$ mismatch) independently and in parallel across
all distances. Each entry passes a comparator ($d_i\!\le\!13$): on pass,
the gate computes $s_i = 255-d_i$ (PASS $\to$ voltage score);
on fail, the output is forced to zero (FAIL $\to$ 0).
This unified voltage representation--higher value means a better
match, is propagated through all three pipeline stages, enabling
CA1 to sort over a single comparable score space.

\textbf{Pipelined bitonic ranking.}
A \emph{pipelined bitonic sorting network}~\cite{mueller2012sortnet}
ranks the similarity scores and selects the Top-1 match. The network
uses $O(\log^2\!D)$ registered compare--exchange stages for $D$ entries,
with fixed latencies of 23 cycles for the 64-entry event bank and
12 cycles for the 16-entry context bank. Ranking latency is independent
of the query and the number of entries that pass the threshold.

\textbf{Synchronizer (done-latch pair).}
The context and event sorters finish after 12 and 23 cycles. Completion
flags retain each result until both are available, then assert output
valid for one cycle. This avoids requiring their non-overlapping done
pulses to coincide.

\section{CA3 Stage: Sequence Prediction}
\label{sec:ca3}

\begin{figure*}[!t]
\centering
\adjustbox{max width=\textwidth, max totalheight=0.42\textheight, keepaspectratio}{\input{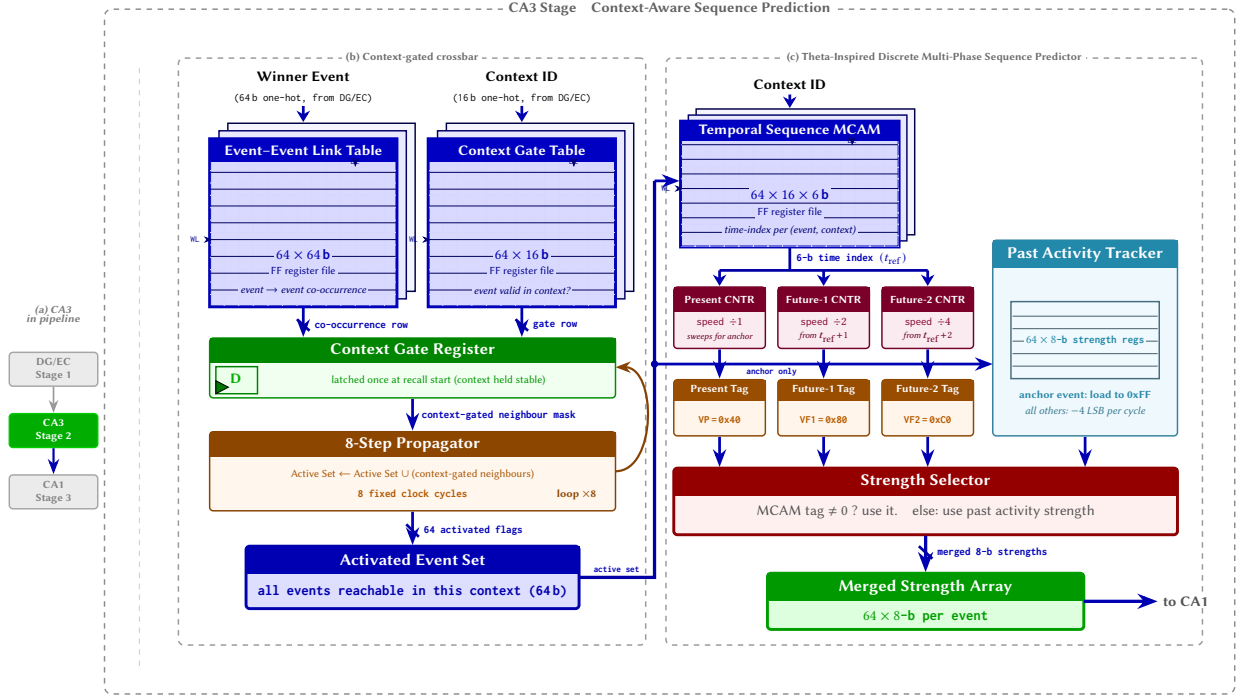}}

\caption{CA3 Stage.
  \textbf{(a)}~Locator.
  \textbf{(b)}~Context-gated crossbar: 64$\times$64 association memory
  and 64$\times$16 gate memory; gate mask registered once per recall
  breaks the feedback cone; active events propagate over 8 fixed hops.
  \textbf{(c)}~3-phase predictor: three counters at $\div$1, $\div$2,
  $\div$4 sweep the 64$\times$16$\times$6-bit MCAM, tagging VP/VF1/VF2.
  Past tracker decays non-anchor events; voltage selector prioritizes
  MCAM tags. Output: 64-element merged voltage array to CA1.}
\Description{Block diagram of the CA3 stage showing the context-gated
  crossbar, the three-phase counter predictor, the past activity tracker
  and the voltage selector.}
\label{fig:ca3_arch}
\end{figure*}

Given the DG/EC event and context one-hot outputs, CA3 performs
context-gated propagation (Fig.~\ref{fig:ca3_arch}b), then tags the present
event and two predicted successors and merges their scores with the
decaying past trace (Fig.~\ref{fig:ca3_arch}c).

\subsection{Context-Gated Crossbar (Fig.~\ref{fig:ca3_arch}b)}
\label{sec:ca3-crossbar}

\textbf{Event--Event Link Table and Context Gate Table.}
Two 64-entry register-file banks hold the learned sequence structure.
The association memory ( $64\!\times\!64$\,b) stores a
\texttt{1} at $(i,j)$ when event~$i$ is directly followed by event~$j$ in
\emph{any} trained context. These fabric registers implement the
recurrent heteroassociative crossbar.
The gate memory ( $64\!\times\!16$\,b) stores a \texttt{1}
at $(j,k)$ when event~$j$ is valid inside context~$k$; it enforces
context boundaries so that associations trained in one context cannot
activate events belonging to another.

\textbf{Context Gate Register.}
The context mask is computed from the stable context one-hot and
registered once at recall start. Subsequent hops use the registered
mask, removing gate-memory reads from the recurrent feedback path.
The earlier combined association/gate cone contained
$64\times(64+16)=5{,}120$ AND terms and prevented timing closure at
100\,MHz.

\textbf{8-Step Propagator.}
Starting from the singleton set $\{j^*\}$, the propagator accumulates for
seven fixed clock cycles (\texttt{MAX\_STEPS}\,$=8$; the eighth cycle only
raises \texttt{done}, giving a 9-cycle start-to-done block latency):
\begin{equation*}
\begin{aligned}
  \mathrm{Active} \;\gets\; \mathrm{Active} \;\cup\; \bigl(
    &\texttt{assoc\_mem}\text{ rows for active events,} \\
    &\text{AND-masked by }\texttt{gate\_results} \bigr)
\end{aligned}
\end{equation*}
OR accumulation ensures that all transitively reachable events, up to
seven hops away, are collected in the activated event set forwarded to
panel~(c).

\subsection{Theta-Inspired Multi-Phase Sequence Predictor (Fig.~\ref{fig:ca3_arch}c)}
\label{sec:ca3-predictor}

\textbf{Temporal Sequence MCAM (Multi-Port CAM).}
A $64\!\times\!16\!\times\!6$-bit register file stores the sequence position
$t \in [0,63]$ of every (event, context) pair loaded during memorisation.
On each recall, CNTR-1 sweeps this memory to locate $t_\mathrm{ref}$,
the position of the anchor event $j^*$ in the current context $k^*$.

\textbf{3-Phase Counter Predictor.}
CNTR-1 counts from 0; CNTR-2 and CNTR-3 then start at $t_\mathrm{ref}$:
\begin{itemize}[noitemsep, topsep=2pt]
  \item \textbf{Present Ctr} ($\div 1$): matches $\mathrm{MCAM}[j^*][k^*]$,
        latching $t_\mathrm{ref}$; tags $j^*$ \textbf{VP}\,=\,0x40.
  \item \textbf{Future-1 Ctr} ($\div 2$): accepts only an activated event
        at $t_\mathrm{ref}{+}1$; tags \textbf{VF1}\,=\,0x80.
  \item \textbf{Future-2 Ctr} ($\div 4$): accepts only
        $t_\mathrm{ref}{+}2$; tags \textbf{VF2}\,=\,0xC0.
\end{itemize}
The offset, not the rate, selects the event; the $\div 2$ and $\div 4$
rates only stagger when the tags are issued, modelling theta-phase
staggering. A timeout at $t_\mathrm{ref}{+}25$ ends an unmatched sweep
(50 cycles at a sequence terminus), so the pipeline cannot stall.

\textbf{Past Activity Tracker.}
Independently of the MCAM predictor, a 64-entry register array
maintains a decaying trace of the past anchor events.
At each recall start the anchor $j^*$ is charged to 0xFF; all
other registers lose 4\,LSB every clock cycle.
Crucially, \emph{only the anchor charges the tracker}: permitting
predicted future events to charge it would corrupt CA1 temporal
ordering on subsequent recalls by making predicted events appear
as past ones.

\textbf{Voltage Selector.}
For each of the 64 events, the selector applies a simple priority rule:
if a non-zero MCAM voltage tag (VP, VF1, or VF2) is available it takes
precedence over the decaying past voltage; otherwise the fading past
trace passes through.
The resulting $64\!\times\!8$-bit merged voltage array encodes a full
past-present-future window and is forwarded to CA1 for temporal
ordering.

\section{CA1 Stage: Temporal Integration}
\label{sec:ca1}

\begin{figure*}[!t]
\centering
\adjustbox{max width=\textwidth, max totalheight=0.42\textheight, keepaspectratio}{\input{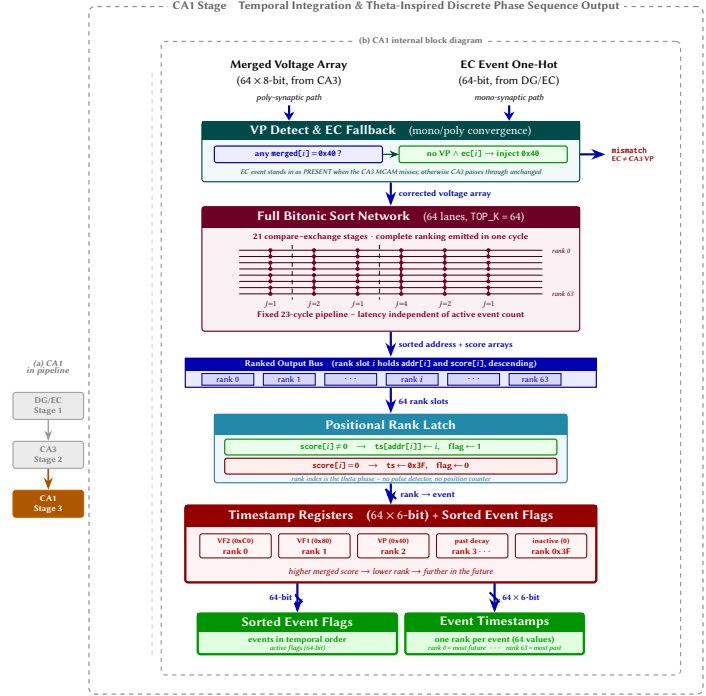}}
\caption{CA1 Stage.
  \textbf{(a)}~Locator.
  \textbf{(b)}~The two EC paths converge: the CA3 merged voltage array
  (poly-synaptic) and the DG/EC event one-hot (mono-synaptic) meet at
  the VP-detect and fallback block, which also raises a
  \texttt{mismatch} flag when the two disagree.
  A full bitonic sorting network (TOP\_K\,=\,64) then ranks all events
  in a fixed 23 cycles: VF2 (0xC0) first, VF1 (0x80), VP (0x40), then
  past residuals; voltage-0 events are never placed.
  The positional rank latch writes each rank index straight into that
  event's timestamp register;
  inactive events retain 0x3F (6-bit sentinel = max).}
\Description{Block diagram of the CA1 stage showing the mono- and
  poly-synaptic inputs converging at a VP-detect and fallback block, a
  full bitonic sorting network, a positional rank latch and the
  per-event timestamp registers.}
\label{fig:ca1_arch}
\end{figure*}

CA1 translates the 64-element merged voltage array, where VP\,=\,0x40
marks the present event, VF1\,=\,0x80 marks near-future, VF2\,=\,0xC0
marks far-future, and decaying values mark past events--into an explicit
ordered sequence with a monotonically-assigned integer timestamp per event.
It is also where the two entorhinal paths converge.
Figure~\ref{fig:ca1_arch} details the sub-blocks.

\subsection{Mono/Poly Convergence and EC Fallback
  (Fig.~\ref{fig:ca1_arch}b, top)}
\label{sec:ca1-ec}

CA1 receives the same EC input twice--poly-synaptically
through DG and CA3, and mono-synaptically straight from EC--and
HippocamFLEX keeps both: the 64-element merged voltage array from CA3,
and the DG/EC top-1 event one-hot tapped before CA3 sees it.
A VP detector scans the merged array for the present tag
(\texttt{merged}$[i]=$\,0x40). 
When CA3 produces no VP tag, CA1 injects the EC-identified event at VP.
When both paths name a present event and disagree, the stage raises a
\texttt{mismatch} flag, making the divergence observable instead of
letting the CA3 answer silently win.

\subsection{Full Bitonic Sorting Network (Fig.~\ref{fig:ca1_arch}b, middle)}
\label{sec:ca1-sorter}

CA1 reuses the DG/EC network with \texttt{TOP\_K}\,=\,64, so the whole
ranking is kept rather than just the winner.
Events land in descending voltage order: VF2 (0xC0\,=\,192) first,
VF1 (0x80\,=\,128) next, VP (0x40\,=\,64) after that, then any
decaying past residuals in order of their remaining charge;
inactive events with voltage\,=\,0 are never placed.
The 64-lane sorting network has 21 compare--exchange stages and a fixed
23-cycle latency, independent of the number of active events.

\subsection{Positional Rank Latch and Timestamp Registers
  (Fig.~\ref{fig:ca1_arch}b, lower)}
\label{sec:ca1-ts}

Because the network emits the complete ordering in one cycle, the rank
index represents position in the theta-inspired sequence; no additional
sequencing logic is needed. For each rank slot~$i$, if the accompanying score is
non-zero the latch writes $i$ into the timestamp register of the event
named by \texttt{addr}$[i]$ and sets that event's flag; if the score is
zero the register takes the sentinel 0x3F and the flag stays clear.
Monotonicity is therefore structurally embedded: rank~0 is the
highest merged voltage (most future), and ranks increase into the past
by construction.

All timestamp registers capture their assigned ranks when the sorter
completes. Their values therefore encode positions in the ranked output,
rather than elapsed clock cycles during the ordering operation.

Together the \textbf{Sorted Event Flags} (64-bit active-event mask) and
\textbf{Event Timestamps} (64 values, with rank~0 for the most future
event and sentinel 0x3F for inactive events) form the theta
inspired-sequence output.

\section{Implementation}
\label{sec:implementation}

\subsection{System-Level Integration}
\label{sec:sysint}

\begin{figure}[!t]
\centering
\resizebox{\columnwidth}{!}{\begin{tikzpicture}[
  font=\rmfamily\small,
  >=Stealth,
  ps/.style     = {draw=blue!55!black,  fill=blue!6,   thick, rounded corners=5pt,
                   minimum width=11.8cm, minimum height=1.2cm, align=center},
  dma/.style    = {draw=teal!60!black,  fill=teal!7,   thick, rounded corners=4pt,
                   minimum width=2.9cm,  minimum height=1.6cm, align=center},
  hip/.style    = {draw=violet!60!black, fill=violet!7, thick, rounded corners=4pt,
                   minimum width=3.8cm,  minimum height=1.8cm, align=center},
  gpio/.style   = {draw=gray!60,        fill=gray!6,   thick, rounded corners=4pt,
                   minimum width=2.7cm,  minimum height=1.6cm, align=center},
  ddr/.style    = {draw=gray!60,        fill=gray!6,   thick, rounded corners=4pt,
                   minimum width=3.5cm,  minimum height=1.1cm, align=center},
  ctrl/.style   = {->, thick, blue!65!black},
  hp0/.style    = {<->, thick, blue!65!black, dashed},
  mm2s/.style   = {->, thick, orange!80!black},
  s2mm/.style   = {->, thick, teal!70!black},
  rst/.style    = {->, thick, gray!70},
  irq/.style    = {->, thick, red!65!black},
]

\node[ps]  (PS)  at (0, 0)
  {\textbf{Zynq MPSoC PS · Cortex-A53}\\[-2pt]
   {\footnotesize HPM0\_FPD (master) · HP0\_FPD (slave) · pl\_ps\_irq0[2:0] · FCLK\_CLK0 100\,MHz}};

\node[dma] (DMA) at (-3.8, -2.8)
  {\textbf{AXI DMA v7.1}\\[-2pt]
   {\footnotesize simple mode · 32-bit}\\[-2pt]
   {\footnotesize\ttfamily 0xA000\_0000}};

\node[hip] (HIP) at ( 0.4, -2.7)
  {\textbf{HippocamFLEX}\\[-2pt]
   {\footnotesize AXIS wrapper · PL core}\\[-2pt]
   {\footnotesize 3-stage pipeline · 100\,MHz}};

\node[gpio](GPIO) at ( 4.5, -2.8)
  {\textbf{AXI GPIO v2.0}\\[-2pt]
   {\footnotesize 8-bit output}\\[-2pt]
   {\footnotesize\ttfamily 0xA001\_0000}};

\node[ddr] (DDR) at (-1.5, -5.2)
  {\textbf{DDR · CMA}\\[-2pt]
   {\footnotesize\ttfamily 0x6BA00000 / 0x6BA01000}};

\draw[ctrl] (PS.south -| DMA.north) -- (DMA.north)
  node[midway, right, font=\rmfamily\footnotesize, blue!65!black]
  {AXI ctrl\,(SC0)};

\draw[ctrl] (PS.south -| GPIO.north) -- (GPIO.north);

\draw[mm2s] (DMA.east |- HIP.west) ++(0, 0.18)
  -- node[above, font=\rmfamily\footnotesize, orange!80!black] {MM2S}
  (HIP.west |- DMA.east) ++(0, 0.18);

\draw[s2mm] (HIP.west |- DMA.east) ++(0,-0.18)
  -- node[below, font=\rmfamily\footnotesize, teal!70!black] {S2MM}
  (DMA.east |- HIP.west) ++(0,-0.18);

\draw[rst] (GPIO.west |- HIP.east) --
  node[above, font=\rmfamily\footnotesize, gray!70] {sw\_rst\_n[0]}
  (HIP.east |- GPIO.west);

\draw[hp0] (DMA.south) |- (DDR.west)
  node[near start, left, font=\rmfamily\footnotesize, blue!65!black] {HP0\,(SC1)};

\coordinate (irqmid) at ($(HIP.north east)!0.5!(GPIO.north west)$);
\draw[irq] (irqmid |- PS.south) -- (PS.south -| irqmid)
  node[midway, right, font=\rmfamily\footnotesize, red!65!black]
  {IRQ[2:0]};

\begin{scope}[shift={(-5.8,-6.5)}]
  \draw[ctrl]         (0,0)   -- ++(0.6,0)
    node[right, font=\rmfamily\scriptsize] {AXI ctrl};
  \draw[hp0]          (2.0,0) -- ++(0.6,0)
    node[right, font=\rmfamily\scriptsize] {HP0 data};
  \draw[mm2s]         (4.0,0) -- ++(0.6,0)
    node[right, font=\rmfamily\scriptsize] {MM2S (cmd)};
  \draw[s2mm]         (6.2,0) -- ++(0.6,0)
    node[right, font=\rmfamily\scriptsize] {S2MM (result)};
  \draw[rst]          (8.6,0) -- ++(0.6,0)
    node[right, font=\rmfamily\scriptsize] {sw\_rst\_n};
  \draw[irq]          (10.5,0) -- ++(0.6,0)
    node[right, font=\rmfamily\scriptsize] {IRQ[2:0]};
\end{scope}
\end{tikzpicture}}
\caption{Vivado block design. HPM0\_FPD$\to$SC0$\to$DMA/GPIO;
  DMA data via SC1 and HP0\_FPD; GPIO bit[0]$\to$\texttt{sw\_rst\_n};
  three IRQ lines.}
\Description{Vivado block design diagram showing the Zynq PS connected
  through SmartConnect to AXI DMA and AXI GPIO, with three interrupt
  lines aggregated by an xlconcat block.}
\label{fig:bd}
\end{figure}
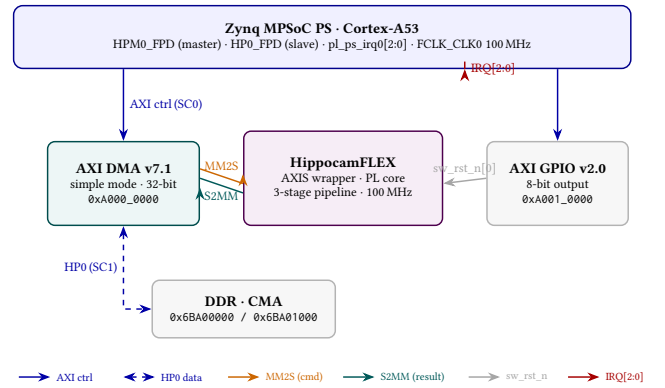

Figure~\ref{fig:bd} shows the four-IP block design on a single
100\,MHz clock domain (\texttt{pl\_clk0}).

\textbf{Address map.}
HPM0\_FPD drives SmartConnect~0 to AXI DMA (0xA000\_0000, 64\,KB)
and AXI GPIO (0xA001\_0000, 64\,KB).
DMA MM2S/S2MM buses reach DDR via SmartConnect~1 and HP0\_FPD; two
CMA buffers are fixed at 0x6BA00000 (TX) and 0x6BA01000 (RX).

\textbf{Reset and interrupts.}
\texttt{pl\_resetn0} fans out to all four peripherals; GPIO bit[0]
drives \texttt{sw\_rst\_n} for firmware warm-reset.
\texttt{xlconcat[2:0]} aggregates \texttt{mm2s\_intr},
\texttt{s2mm\_intr}, and \texttt{irq\_result\_ready} into
\texttt{pl\_ps\_irq0[2:0]}; the wrapper asserts
\texttt{irq\_result\_ready} as a one-cycle pulse on result capture.

Table~\ref{tab:cmds} defines the seven-opcode AXI4-Stream ISA.

\begin{table*}[t]
  \caption{HippocamFLEX AXI4-Stream ISA. \textit{Beat} = one 32-bit word;
    opcode in bits [31:28] of Beat~0; HV[127:0] sent LSB-first over 4 data-beats.}
  \label{tab:cmds}
  \small
  \begin{tabular}{llrrl}
    \toprule
    \textbf{Cmd} & \textbf{Opcode} & \textbf{Beats} & \textbf{Bytes}
      & \textbf{Beat~0 encoding} \\
    \midrule
    Memorise Context HV    & 0x1 & 5 & 20 &
      {[31:28]=0x1, [6:0]=addr; Beats 1--4: context HV[127:0]} \\
    Memorise Event HV      & 0x2 & 5 & 20 &
      {[31:28]=0x2, [6:0]=addr; Beats 1--4: event HV[127:0]} \\
    Link Events (Crossbar) & 0x3 & 1 &  4 &
      {[31:28]=0x3, [20:14]=from\_event, [13:7]=to\_event, [0]=1} \\
    Assign Context Gate    & 0x4 & 1 &  4 &
      {[31:28]=0x4, [20:14]=event, [13:7]=context, [0]=1} \\
    Set Timing (MCAM)      & 0x5 & 1 &  4 &
      {[31:28]=0x5, [20:14]=event, [13:7]=context, [5:0]=code} \\
    Recall (Query)         & 0x6 & 9 & 36 &
      {[31:28]=0x6; Beats 1--4: ctx\_hv; Beats 5--8: evt\_hv} \\
    Status Query           & 0x7 & 1 &  4 &
      {[31:28]=0x7 (no payload)} \\
    \bottomrule
  \end{tabular}
\end{table*}

\subsection{Design Parameters and Memory}

Tables~\ref{tab:params} and~\ref{tab:mem} list parameters and memory.

\begin{table}[t]
  \caption{Key design parameters.}
  \label{tab:params}
  \small
  \resizebox{\columnwidth}{!}{%
  \begin{tabular}{lrl}
    \toprule
    \textbf{Parameter} & \textbf{Value} & \textbf{Biological analogue} \\
    \midrule
    HV\_WIDTH      & 128\,b         & EC hypervector dimensionality \\
    NUM\_CONTEXTS  & 16             & Max distinct contexts \\
    NUM\_EVENTS    & 64             & Max distinct events \\
    MAX\_MISMATCH  & 13\,b          & 90\% similarity threshold (DG) \\
    MAX\_STEPS     & 8              & CA3 recurrent depth \\
    CNTR\_WIDTH    & 6\,b           & Temporal code range (0--63) \\
    VOLTAGE\_WIDTH & 8\,b           & Digital ML voltage (0x00--0xFF) \\
    DECAY\_STEP    & 4\,LSB         & Capacitor discharge rate \\
    VP/VF1/VF2     & 0x40/0x80/0xC0 & Present/Future-1/Future-2 \\
    Clock          & 100\,MHz       & PS FCLK\_CLK0 \\
    \bottomrule
  \end{tabular}%
  }
\end{table}

\begin{table}[t]
  \caption{On-chip memory inventory.}
  \label{tab:mem}
  \small
  \resizebox{\columnwidth}{!}{%
  \begin{tabular}{lll}
    \toprule
    \textbf{Memory} & \textbf{Dimensions} & \textbf{Type} \\
    \midrule
    Context Proximity-CAM   & $16\times128$\,b       & Register file \\
    Event Proximity-CAM     & $64\times128$\,b       & Register file \\
    Crossbar assoc.\ memory & $64\times64$\,b        & Register file \\
    Crossbar gate memory    & $64\times16$\,b        & Register file \\
    MCAM temporal memory    & $64\times16\times6$\,b & Register file \\
    Past-voltage registers  & $64\times8$\,b         & FF array \\
    CA1 timestamp registers & $64\times6$\,b         & FF array \\
    \bottomrule
  \end{tabular}%
  }
\end{table}

\subsection{Design, Synthesis and Verification}

All memories map to FF arrays (Table~\ref{tab:mem})
for three reasons: the Proximity-CAM search requires results in one
registered stage (BRAM adds a mandatory second); the crossbar requires
1-bit write granularity BRAM ports cannot provide; and the MCAM uses two
simultaneous variable indices Vivado will not infer onto dual-port BRAM.
The datapath uses \texttt{\$countones}$(q\oplus m_i)$ for
8-bit Hamming distances, clock-enable divisors ($\div$2, $\div$4) for
temporal tagging, bitonic sorting networks for Top-1 selection and full
ranking, and a linear 4\,LSB/cycle decrement for the past-event trace.

Synthesis targets Vivado 2023 on XCZU9EG (ZCU102). 
Five memorisation-control FFs needed false-path annotations, and the
high-fanout activated register required
\texttt{MAX\_FANOUT=64}.
Because every memory is a flip-flop array, the design uses no BRAM,
URAM or DSP (Table~\ref{tab:ppa}), leaving the entire hard-block fabric
free for a co-resident accelerator.
Three testbench levels cover the design: a \emph{unit} bench for Proximity-CAM
Top-1 selection; an \emph{integration} bench for CA3 voltage tags and CA1
timestamp monotonicity on the ZABCD sequence; and an \emph{AXI wrapper}
bench exercising all seven opcodes, back-pressure, and the full 35-beat
response.
All three pass under both behavioural and post-implementation simulation.
Table~\ref{tab:perf} summarises the latency profile. With fixed-latency
sorting networks the two ranking stages no longer dominate, and CA3's
anchor sweep--whose length is set by the stored temporal code
$t_\mathrm{ref}$--becomes the critical term.

\begin{table}[t!]
  \caption{Pipeline latency and performance summary at 100\,MHz.
    $^{\dagger}$MCAM sweep is data-dependent: 18 cycles at
    $t_\mathrm{ref}{=}3$ with both predictions found, 59 at a
    sequence terminus.}
  \label{tab:perf}
  \small
  \begin{tabular}{lrl}
    \toprule
    \textbf{Metric} & \textbf{Value} & \textbf{Notes} \\
    \midrule
    DG/EC latency   & 25 cyc (0.25\,\textmu s) &
      1 popcount + 23-cyc bitonic \\
    CA3 latency     & 27 cyc (0.27\,\textmu s) &
      9 crossbar + 18 MCAM$^{\dagger}$ \\
    CA1 latency     & 24 cyc (0.24\,\textmu s) &
      21-stage bitonic, TOP\_K\,=\,64 \\
    \textbf{Total} & \textbf{76 cyc (0.76\,\textmu s)} &
      Seq.\ DG/EC$\to$CA3$\to$CA1 \\

    \bottomrule
  \end{tabular}
\end{table}

\begin{figure*}[!t]
\centering
\includegraphics[width=0.99\textwidth,height=5.3in,keepaspectratio]{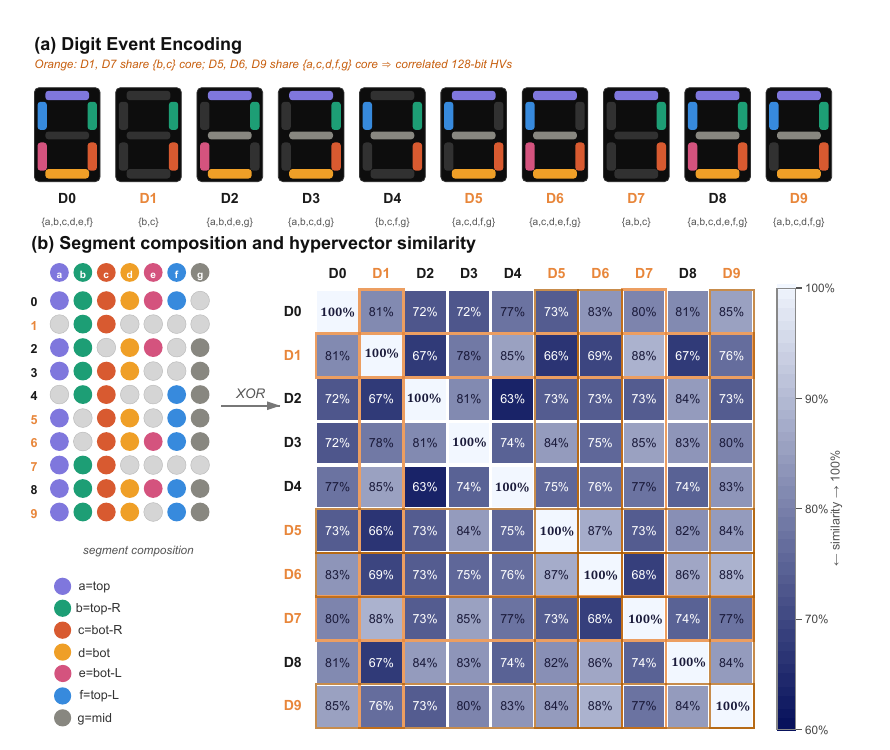}
\caption{7-segment benchmark.
  \textbf{(a)}~D0--D9 mapped to active segments; orange digits share
  correlated segment cores (D1/D7 share \{b,c\}; D5/D6/D9 share
  \{a,c,d,f,g\}), stressing the DG/EC pattern-separation stage.
  \textbf{(b)}~Segment composition matrix and pairwise HV similarity
  ($1{-}d_\mathrm{Hamming}/128$). Off-diagonal values range 63--88\%;
  diagonal (self-similarity) shown as filled cells.}
\Description{Two-panel figure showing seven-segment digit encodings and
  a pairwise hypervector similarity matrix.}
\label{fig:7seg}
\end{figure*}

\subsection{Resource Scaling}
\label{sec:scaling}

We swept event capacity $E$ to measure its resource and latency costs. We 
synthesised $E\in\{32,64,128\}$ with everything else fixed
(Table~\ref{tab:ppa}).

The resource model has three terms:
\begin{equation}
  \underbrace{E^{2}}_{\text{crossbar}} \;+\;
  \underbrace{E\log_2^{2}E}_{\text{sorters}} \;+\;
  \underbrace{E}_{\text{memories}},
  \label{eq:area}
\end{equation}
and only the crossbar is quadratic: associate memory holds one bit per
ordered pair of events. Latency touches $E$ only through the two ranking
stages,
\begin{equation}
  \mathrm{cycles}(E) \;\approx\; \log_2\!E\,(\log_2\!E + 1) + 31 + t_\mathrm{ref} .
  \label{eq:latency}
\end{equation}
Quadrupling capacity from 32 to 128 events costs $7.7\times$ the LUTs but
only 41\% more cycles (64 to 90 at $t_\mathrm{ref}{=}3$). 

Only the $t_\mathrm{ref}$ term is data-dependent. CNTR-1 sweeps from zero,
so locating the anchor costs $t_\mathrm{ref}{+}1$ cycles; the two future
counters then resolve in 9 more when both predictions are found, but at a
sequence terminus no event carries $t_\mathrm{ref}{+}1$ or
$t_\mathrm{ref}{+}2$ and the sweep instead runs to its timeout, 50 cycles.
Recall at $E{=}64$ therefore spans 76 cycles in the common case to 117 at a
terminus. Reducing the MCAM sweep would address this variable latency.

Two ceilings meet at $E{=}128$: the AXI4-Stream header packs the event
index into seven bits, and the design already fills 80\% of the device.
Sparse or banked association storage is a candidate for reducing
the capacity cost.

\begin{table}[t]
  \caption{Post-place resources versus event capacity on XCZU9EG
    (274{,}080 LUT, 548{,}160 FF). All other parameters fixed;
    BRAM, URAM and DSP are zero at every point.}
  \label{tab:ppa}
  \small
  \begin{tabular}{rrrrrr}
    \toprule
    \textbf{Events} & \textbf{LUT} & \textbf{\%} & \textbf{FF} & \textbf{\%}
      & \textbf{Cycles} \\
    \midrule
     32 &  28{,}661 & 10.5 &  25{,}214 &  4.6 &  64 \\
     64 &  76{,}056 & 27.8 &  58{,}061 & 10.6 &  76 \\
    128 & 220{,}287 & 80.4 & 146{,}132 & 26.7 &  90 \\
    \bottomrule
  \end{tabular}
\end{table}

\subsection{Timing Closure}
\label{sec:fmax}
Both deployable configurations were implemented against an 8.333\,ns
(120\,MHz) constraint, tighter than the 100\,MHz deployment clock, and
close timing with no failing setup or hold endpoints. Worst setup slack
(WNS) is $+0.316$\,ns at $32\!\times\!16$ and $+0.628$\,ns at
$64\!\times\!16$, giving $F_{\max}=1000/(T-\mathrm{WNS})$ of 124.7 and
129.8\,MHz, i.e., ${\approx}25\%$ margin at 100\,MHz. The larger design's
slightly higher $F_{\max}$ reflects place-and-route seed variance, not a
trend.

\section{Results}
\label{sec:results}

\subsection{Simulation Results}

All three testbenches (\S\ref{sec:implementation}) pass behavioural and
post-implementation simulation in Vivado 2023 with no assertion failures;
the unit bench also confirms Proximity-CAM Top-1 within the $\le$13-bit
window for every trained HV.

\subsection{7-Segment Benchmark on Silicon}

\subsubsection{Benchmark Design}

A hippocampal benchmark requires events that are ambiguous alone;
correct recall depends on context. Ten digit patterns (D0--D9) are
defined over eight colour contexts; the on-silicon suite exercises seven
digits (D1--D7) across five of them (Red, Green, Blue, Magenta, White).
Each digit HV is the OR of fixed random
128-bit segment bases; context HVs are composites of RGB channel bases.
Correlated digits (e.g., D3, D5, D7 all share the $\{a,c\}$ core)
deliberately stress DG pattern separation.

\subsubsection{On-Silicon Results}

The statically-linked AArch64 driver loads all contexts and events,
writes crossbar and MCAM entries, and runs eight recall cases
(Table~\ref{tab:results}); all 36 scoreboard checks pass.
T1--T5 validate VP/VF1/VF2 assignment and monotonic CA1 ordering
across five clean contexts.
T6 (Blue/D3 mid-sequence) confirms the sliding window: D2 receives
voltage 0x00 and CA1 sentinel 0x3F, showing the prediction window
advanced without hallucinating a past event as future.

\begin{figure}[!htb]
  \centering
  \includegraphics[width=\columnwidth,height=3.2in,keepaspectratio]{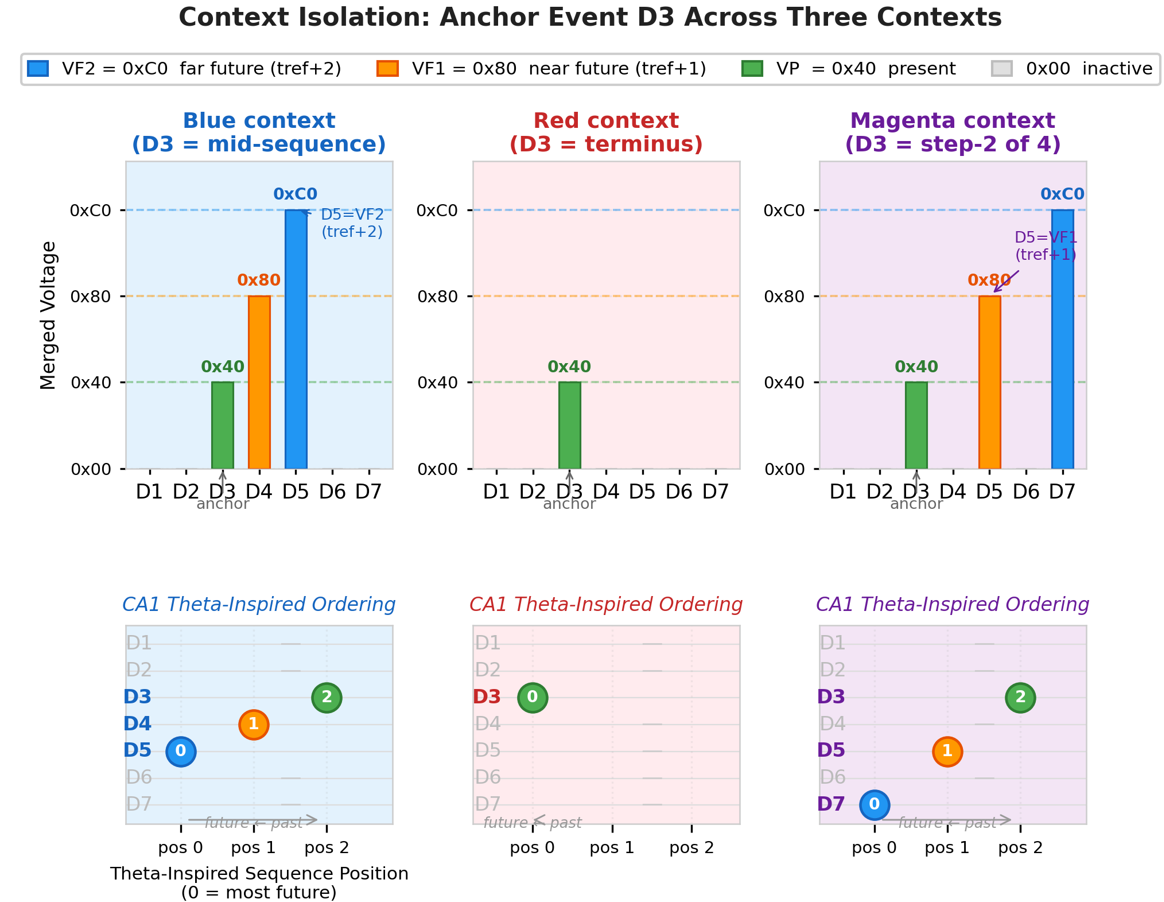}
  \caption{On-silicon context isolation (anchor D3, three contexts).
    \textbf{Top}: merged-voltage profiles; dashed lines at VP/VF1/VF2
    tiers.
    \textbf{Bottom}: CA1 rank timestamps (0\,=\,most future).
    Two distinct mechanisms hold the boundary: under Blue the gate memory
    excludes D7 outright, whereas under Red D5 and D7 are activated by the
    shared edges but receive no tag, because D3 is the sequence terminus
    and Red holds no event at $t_\mathrm{ref}{+}1$ or $t_\mathrm{ref}{+}2$.}
  \Description{Two-panel plot of merged voltage profiles and CA1
    timestamps for anchor D3 across three colour contexts.}
  \label{fig:context_isolation}
\end{figure}

T7a/T7b are the stress tests: context isolation under shared crossbar
edges. In T7a (Blue/D3), D7 must stay inactive despite
D3$\to$D5$\to$D7 existing in Magenta--the gate memory holds the
boundary. In T7b (Red/D3), D3 is terminal; both D4 and D2 show
0x00 voltage and 0x3F timestamps, confirming correct abstention.
Figure~\ref{fig:context_isolation} visualises the three-context
comparison: D5 acquires VF2 under Blue ($t_\mathrm{ref}{+}2{=}5$)
but VF1 under Magenta ($t_\mathrm{ref}{+}1{=}3$), confirming that
MCAM tags are context-relative, not event-identity-relative.

\begin{table}[t]
  \caption{On-silicon memorise-recall. ``CA1 order'':
    ts[F2]$<$ts[F1]$<$ts[VP].
    ``---'' denotes no prediction; ``n.c.'' denotes a tier the test does
    not assert.}
  \label{tab:results}
  \small
  \begin{tabular}{llllll}
    \toprule
    \textbf{Test} & \textbf{Ctx} & \textbf{Anchor}
      & \textbf{VP} & \textbf{VF1/VF2} & \textbf{Check} \\
    \midrule
    T1  & Blue    & D1 & D1 & D2/D3   & CA1 order  \\
    T2  & Red     & D7 & D7 & D5/D3   & CA1 order  \\
    T3  & Green   & D2 & D2 & D4/D6   & CA1 order  \\
    T4  & Magenta & D1 & D1 & D3/D5   & CA1 order  \\
    T5  & White   & D3 & D3 & D5/D7   & CA1 order  \\
    T6  & Blue    & D3 & D3 & D4/D5   & D2 inactive  \\
    T7a & Blue    & D3 & D3 & D4/n.c.\ & D7 inactive  \\
    T7b & Red     & D3 & D3 & ---     & D4,D2 inactive  \\
    \bottomrule
  \end{tabular}
\end{table}
\section{Discussion}
\label{sec:discussion}

\textbf{Context isolation (T7a, T7b)} uses complementary mechanisms:
context gates exclude irrelevant events, while context-specific timing
codes suppress unsupported continuations at sequence termini.
CA1's monotonic timestamps directly realise theta-sequence ordering in
synthesisable logic~\cite{shahbaba2022}, demonstrating temporal
integration within the complete FPGA pipeline.

The main cost is the register-based association memory, which grows
quadratically with event capacity; the bitonic networks keep ranking
latency fixed per configuration, so the MCAM anchor search is the only
data-dependent latency term.

Future work includes offline replay of prioritized event-in-context
sequences into a cortical association graph, probabilistic Top-K recall,
and C3M (Conformal Configurable Compute-in-Memory) implementations for
area and energy studies. These extensions are unevaluated; the verified
digital behavior can provide regression tests.

\section{Conclusion}
\label{sec:conclusion}
HippocamFLEX demonstrates that proximity-based pattern separation,
context-gated recurrent prediction, and theta inspired-sequence temporal
integration can be jointly realised in synthesisable RTL and verified on
physical silicon. All eight on-silicon test cases pass; context
isolation holds under shared crossbar edges, CA1 timestamps are
monotonically ordered, and sequence termini are correctly detected as
prediction absence rather than hallucinated continuations.
In summary, HippocamFLEX is the first work to translate biological
architectural details of hippocampal sub-regions to an FPGA platform,
laying the groundwork for emulating other brain sub-regions on FPGAs.

\begin{acks}
This work was supported in part by the National Science Foundation under
Award Nos.~CCF-2319619, CCF-2319618, and CCF-2319617. This work was also partially
supported by the National Institutes of Health under Award
Nos.~R01-MH115697, R01-DC017687, and R01-CA297869, and by the National
Science Foundation under Award Nos.~CAREER IOS-1150292, BCS-1439267,
and DGE-1839285.
\textbf{\textit{AI disclosure:}} LLMs were used only to enhance the architecture
diagrams; they were \textbf{NOT} used for the ideation, design, results,
data or result figures.
\end{acks}

\FloatBarrier

\balance
\bibliographystyle{ACM-Reference-Format}
\bibliography{bib}

\end{document}